\documentclass[10pt,preprint2]{aastex}
\usepackage{natbib}
\usepackage{layout}
\usepackage[latin1]{inputenc}
\usepackage{amssymb}
\usepackage{lscape}

\shorttitle{ESO VLT Optical Spectroscopy of BL Lac Objects IV - New spectra and properties of the full sample.}
\shortauthors{M. Landoni et al.}

\begin{document}

\title{ESO VLT Optical Spectroscopy of BL Lac Objects IV - New spectra and properties of the full sample.}

\author{M. Landoni}
\affil{Universit\`{a} degli Studi dell' Insubria. Via Valleggio 11, I-22100 Como, Italy.}
\affil{INAF - Osservatorio Astronomico di Brera, Via Emilio Bianchi 46, I-23807 Merate, Italy}
\affil{INFN - Istituto Nazionale di Fisica Nucleare}
\email{marco.landoni@uninsubria.it}
\author{R. Falomo}
\affil{INAF - Osservatorio Astronomico di Padova, Vicolo dell' Osservatorio 5, I-35122 Padova, Italy}
\author{A. Treves}
\affil{Universit\`{a} degli Studi dell' Insubria. Via Valleggio 11, I-22100 Como, Italy.}
\affil{INAF - Istituto Nazionale di Astrofisica and INFN - Istituto Nazionale di Fisica Nucleare}
\author{B. Sbarufatti}
\affil{INAF - Osservatorio Astronomico di Brera, Via Emilio Bianchi 46, I-23807 Merate, Italy}
\author{M. Barattini}
\affil{Universit\`{a} degli Studi dell' Insubria. Via Valleggio 11, I-22100 Como, Italy.}
\author{R. Decarli}
\affil{Max-Planck-Institut fur Astronomie, Konigstuhl 17, 69117 Heidelberg, Germany}
\author{J. Kotilainen}
\affil{Finnish Centre for Astronomy with ESO (FINCA) - University of Turku, V\"ais\"al\"antie
 20, FI-21500 Piikki\"o, Finland}

\begin{abstract}
We present the last chapter of a spectroscopy program aimed at deriving the redshift or a lower limit to the redshift of BL Lac objects using medium resolution spectroscopy.
Here we report new spectra for 33 BL Lac object candidates obtained in 2008-2009 confirming the BL Lac nature of 25 sources and for 5 objects we obtained new redshifts. These new observations are combined with our previous data in order to construct a homogeneous sample of $\sim$ 70 BL Lacs with high quality spectroscopy. All these spectra can be accessed at the website \textit{http://www.oapd.inaf.it/zbllac/}. The average spectrum, beaming properties of the full sample, discussion on intervening systems and future perspectives  are addressed.
\end{abstract}


\section{Introduction}
BL Lac objects are extragalactic sources for which the redshift $z$ cannot be easily derived. This is due to the combination of intrinsically weak emission lines and strongly beamed nuclear non thermal emission.
The determination of $z$ is obviously important in order to properly characterize their physical properties. While other distance indicators
were introduced like the photometry of the host galaxy (imaging redshift, see e.g.
\citet{sba05}), or TeV $\gamma$-ray absorption by the EBL (e.g. \citet{prand10}), the direct spectroscopic line observation remains the only way to produce a robust measure of their distance.
The determination of $z$ is becoming even more compelling by the recent development of $\gamma$-ray astronomy, which demonstrated that the extragalactic $\gamma$-ray sky is dominated by this class of objects. 
\\
\\
In order to improve the quality (in terms of throughput, spectral resolution and S/N ratio) of the spectrum, we have carried out systematic campaigns at European Southern Observatory Very Large Telescope (VLT), aimed to observe $\sim$ 90 BL Lac candidates of unknown redshift. Results on the first three runs were presented in \cite{sba05a, sba06_02,sba09}, Paper I, II and III respectively , where the criteria of selection of the sample (see also Section \ref{subsec:sample}), details on the used spectrometer, and the reduction and analysis procedure are given.  
\\
\\
In this paper we report on observations of 33 new BL Lac candidates that complete our observational campaign and for 25 of them we confirm their BL Lac nature.
In Section \ref{sec:observations} we summarize the observations and data reduction process and we show the results in Section \ref{sec:result}. The average spectrum of the class is presented in Section \ref{sec:specmean}, while a discussions of the properties for the full sample is reported in Section \ref{sec:discussion}.
\\
\\
Throughout the paper we adopt the following concordant cosmology: $H_{0} = 70$ km s$^{-1}$ Mpc$^{-1}$, $\Omega_{m} = 0.27$ and $\Omega_{\Lambda} = 0.73$.

\section{Observations and data reduction}
\label{sec:observations}
The spectra were obtained in Service Mode using the FOcal Reducer and a 
low dispersion Spectrograph (FORS2, \citet{seif2000}). The observations took place between April 2008 and June 2008 at VLT-UT1 (Antu) and from September 2008 to February 2009 at VLT-UT2 (Kueyen). The spectrograph was configured with a 300V+I grism with a slit of 2$^{\prime\prime}$ achieving the spectral range of  3800 - 8300 $\textrm{\AA}$ with a dispersion of $\sim$ $3.2 \textrm{\AA}$ /pixel. Detailed informations on the observations of each source are given in Table \ref{tab:journal}. 
The seeing during the observation was in the range 0.65$^{\prime\prime}$-2.50$^{\prime\prime}$ with an average of 1.20$^{\prime\prime}$ (rms $\sim 0.4^{\prime\prime}$). 
\\
\\
Data reduction was executed using IRAF\footnote{1
IRAF (Image Reduction and Analysis Facility) is distributed by the National Optical Astronomy Observatories,
which are operated by the Association of Universities for
Research in Astronomy, Inc., under cooperative agreement
with the National Science Foundation.} adopting standard procedures. In particular, for each target we performed bias subtraction, flat fielding and bad pixels or cosmic rays rejection. For each object we secured 2-6 individual exposures that
were then combined into a single average image. 
The wavelength calibration was achieved using the spectra of a Helium Neon Argon lamp reaching an accuracy of $\sim$ 3 $\textrm{\AA}$ rms.
Even though this program did not require photometric conditions during observations, checking the weather database at Paranal, the sky was clear during most of the time. This enables us to flux calibrate our spectra using standard stars \citep{oke1990}. The flux calibration was   assessed through the relative photometric calibration of the acquisition image and the overall uncertainty is $\Delta m \sim 0.1$ mag (rms).
Finally, each spectrum has been dereddened applying the extinction law described in \cite{card89} assuming the $E_{B-V}$ values computed by \cite{sch98}.

\begin{small}
\begin{deluxetable}{lllllcclll}
\tabletypesize{\scriptsize}

\tablecaption{Journal of the observations}

\tablenum{1}

\label{tab:journal}

\tablehead{\colhead{Object} & \colhead{R.A.(J2000)} & \colhead{Decl.(J2000)} & \colhead{R} & \colhead{Date} & \colhead{Exposure time}  & \colhead{N} & \colhead{S/N}  &\colhead{Seeing} & \colhead{Ref} \\
\colhead {(1)} & \colhead{(2)} & \colhead{(3)} & \colhead{(4)} & \colhead{(5)} & \colhead{(6)} & \colhead{(7)} & \colhead{(8)} & \colhead{(9)} & \colhead{(10)}  \\} 

\startdata
\object{GC 0109+224} & 01 12 09.103 & $+$ 22 44 35.052 & 15.60 & 2008-09-20  & 2040 & 6 & 200 & 2.00 &  V01 \\

\object{PKS 0140-059} & 01 42 40.810 & $-$ 05 43 26.472 & 18.20 & 2008-10-26 & 1362 & 2 & 65 & 1.00 & V06 \\

\object{RBS 0337} & 02 38 29.549 & $-$ 31 16 33.312 & 16.40& 2008-09-30 & 1362 & 2 & 130 & 0.65 &V01 \\

\object{PKS 0302-623} & 03 03 56.678 & $-$ 62 11 40.812 & 17.90 & 2008-09-27 & 1362 & 2 & 150 & 1.40 & S94 \\

\object {RBS 0499} & 03 58 55.903 & $-$ 30 54 02.664 & 17.60 & 2008-09-20 & 1362 & 2 & 40 & 1.00 & V01 \\

\object {PKS 0410-519} & 04 11 41.858 & $-$ 51 49 31.548 & 16.50 & 2008-10-01 & 1362 & 2 & 70 & 1.60 & G04 \\

\object {PKS 0439-299} & 04 41 22.987 & $-$ 29 52 31.296 & 18.00 & 2008-10-25 & 1362 & 2 & 30 & 1.20 & V01 \\

\object {PKS 0443-387} & 04 45 14.474 & $-$ 38 38 34.296 & 17.00 & 2008-10-26 & 1362 & 2 & 140 & 1.70 & G04\\

\object {RBS 0589} & 04 48 40.733 & $-$ 16 32 51.684 & 16.60 & 2008-10-28 & 1362 & 2 & 130 & 0.80 & V01 \\

\object {ZS 0506+056} & 05 09 25.900 & $+$ 05 42 21.000 & 15.10 & 2008-11-02 & 800 & 2 & 200 & 2.00 & V01 \\

\object {PMN J0529-3555} & 05 29 40.800 & $-$ 35 55 41.160 & 18.90 & 2008-10-29 & 1362 & 2 & 30 & 1.30 & G04 \\

\object {PKS 0539-543} & 05 40 50.200  & $-$ 54 17 58.700 & 17.30 & 2008-10-29 & 1362 & 2 & 70 & 1.60 & G04 \\

\object {PKS 0607-605} & 06 08 00.334 & $-$ 60 31 27.732 & 17.60 & 2008-10-29 & 1362 & 2 & 50 & 1.90 & G04 \\

\object {MS-0622.5-5256} & 06 23 42.667 & $-$ 52 58 01.704 & 19.20 & 2008-10-26 & 1362 & 2 & 45 & 2.50 & V01\\

\object {PKS 0735+17} & 07 38 08.100 & $+$ 17 43 03.600 & 15.80 & 2009-01-13 & 2720 & 4 & 160 & 1.30 & V01 \\

\object {OJ-131	} & 08 20 55.562 & $-$ 12 59 35.340 & 16.50 & 2008-04-09 & 1362 & 2 & 100 & 1.60 & V01 \\

\object {RX J1022.7-0112} & 10 22 46.704 & $-$ 01 13 11.280 & 16.90 & 2008-04-09 & 1361 & 2 & 100 & 1.50 & V01 \\

\object {RX J1023.9-4336} & 10 23 56.446 & $-$ 43 36 46.980 & 16.70 & 2008-04-10 & 1361 & 2 & 200 & 1.40 & V01 \\

\object {PKS B1056-113} & 10 59 09.478 & $-$ 11 34 36.516 & 16.20 & 2008-04-18 & 1361 & 2 & 100 & 1.50 & V06 \\

\object {1ES 1106+244} & 11 09 18.245 & $+$ 24 10 45.156 & 18.70 & 2008-04-22 & 2720 & 4 & 40 & 1.30 & V01 \\

\object {RX J1117.0+2014} & 11 17 07.522 & $+$ 20 14 49.632 & 15.10 & 2009-02-16 & 2612 & 4 & 130 & 1.00 & V01 \\

\object {RX J1241.8-1455} & 12 41 49.862 & $-$ 14 56 42.756 & 17.70 & 2008-04-22  & 2720 & 4 & 140 & 0.95 & V01 \\

\object {1ES 1239+069} & 12 41 51.314 & $+$ 06 36 06.840 & 18.50 & 2008-04-17/22 & 2720 & 4 & 50 & 1.00 & V01 \\

\object {4C 12.46} & 13 09 31.862 & $+$ 11 53 50.676 & 18.20 & 2008-04-18 & 1361 & 2 & 70 & 1.00 & V01 \\

\object {OQ-240} & 14 27 03.307 & $+$ 23 47 39.840 & 14.50 & 2008-04-07/17 & 800 & 4 & 170 & 1.00 & V01 \\

\object {OR 103} & 15 04 27.670 & $+$ 10 29 17.520 & 18.90 & 2009-02-16 & 1361 & 2 & 150 & 1.70 & V01 \\

\object {RBS 1478} & 15 16 16.718 & $-$ 15 24 18.000 & 18.50 & 2008-04-02 & 1361 & 2 & 50 & 0.90 & V01 \\

\object {PKS 1815-553} & 18 19 44.678 & $-$ 55 22 06.204 & 21.50 & 2008-04-22 & 1362 & 2 & 10 & 1.30 & S94 \\

\object {OW-080} & 20 50 08.688 & $+$ 04 08 15.792 & 19.00 & 2008-04-22 & 1361 & 2 & 90 & 0.70 & V01 \\

\object {OX-183} & 21 52 27.972 & $+$ 17 34 30.396 & 18.00 & 2008-06-02 & 2722 & 4 & 50 & 0.90 & V01 \\

\object {RBS 1899} & 22 49 13.829 & $-$ 13 00 05.616 & 18.20 & 2008-06-08 & 1361 & 2 & 60 & 1.00 & V01 \\ 

\object {RX J2319.6+1611} & 23 19 44.371 & $+$ 16 12 32.688 & 17.30 & 2008-06-08 & 1361 & 2 & 100 & 1.70 & V01 \\

\object {1ES 2322-409} & 23 24 46.238 & $-$ 40 41 31.128 & 15.70 & 2008-05-28 & 1361 & 2 & 60 & 0.90 & V01 \\
\enddata

\tablecomments{Description of columns: 
(1) Object name;     
(2) Right ascension (J2000); 
(3) Declination (J2000);
(4) R apparent magnitude from catalog; 
(5) Date of observation; 
(6) Exposure time (s); 
(7) Number of spectra collected; 
(8) Mean signal-to-noise ratio; 
(9) Mean seeing during observation
(10) References: \textbf{G04}: \cite{giommi2004};  \textbf{S94}: \cite{stickel94}; \textbf{V01}: \cite{veron01};  
}

\

\end{deluxetable}
\end{small}

\section{Results}

\label{sec:result}
From the spectroscopic point of view we consider as bona-fide BL Lac the sources that either have a featureless spectrum or in which their intrinsic features have equivalent width (EW) $\lesssim$ 5 \AA. Based on this classification we confirm the BL Lac nature for 25 objects. 
Their spectra are reported in Figure \ref{fig:spec} while confirmed non-BL Lac spectra are shown in Figure \ref{fig:spectraQSO}. All BL Lac spectra can be accessed at the website \textit{http://www.oapd.inaf.it/zbllac/}
In particular, 20 sources have no intrinsic lines, 3 of them show weak absorption features arising from their host galaxy and other 3 show weak narrow emission lines. For only one object (OX 183) a weak broad emission line is apparent.   
We identify spectral features arising from the source or from intervening systems with the atomic species (if applicable) while the Diffused Interstellar Bands are marked with the label DIB. Telluric absorption are indicated by the symbol $\oplus$.
The identification of the lines, their equivalent width, full width at half maximum and redshift estimation are given in Table \ref{tab:results} for BL Lac objects and in Table \ref{tab:resultsqso} for the other sources.
We also report in Appendix A short comments on interesting individual sources .

\subsection{The continuum emission}
In the optical region the emission of BL Lac objects is characterised by a superposition of a non-thermal component arising by the accreting nucleus and a thermal one that originates from its host galaxy.
In particular, the emission from the nucleus can be described by a power law model ($F_{\lambda} \propto \lambda^{-\alpha}$) while the galaxy contribution can be modeled by an elliptical galaxy template spectrum with $M_{R}^{host} = -22.90 \pm 0.50$ \citep{sba05}.
For the objects that do not present host galaxy absorption features we fitted the continuum  adopting a single power law, while for the BL Lac where galactic absorption are detected we performed the fit by adding the host galaxy template contribution, following procedure described in Paper II and III. Resulting power law indexes are reported in Table \ref{tab:results}. The mean value is $\alpha_{mean} = 0.87 \pm 0.35 $ which is consistent with result for power law indexes obtained in Paper I, II and III ($\alpha_{mean} = 0.90 \pm 0.40 $).

\subsection{Line detection and EW$_{min}$}
We measured the minimum detectable equivalent width (EW$_{min}$) for each spectrum adopting the recipe described in Paper I. We considered as line canditates all the spectral features revealed by an automatic procedure selecting EW  above EW$_{min}$. Each candidate has been carefully inspected for the consequent feature identification or rejection. 
Spectral parameters, line properties and relative identification and EW$_{min}$ for the BL Lac objects investigated in this paper are reported in Table \ref{tab:results}.

\subsection{Redshift lower limits}
\label{sec:specdec}
Despite high quality of the available instrumentation in terms of throughput and the high signal-to-noise ratio reached, 20 objects still show a featureless continuum. For these sources, we applied the procedure fully illustrated in  
Paper II and III in order to measure a lower limit on the redshift of each source.
Briefly, assuming that the BL Lac host galaxy is a giant elliptical with $M_{R}^{host} = -22.90 \pm 0.50$ if one measures the EW$_{min}$ and the nucleus apparent magnitude it is possibile to infer a lower limit on the $z$ of the source using Equation (Paper III)
\begin{equation}
\label{eq:1}
EW_{obs} = \frac{(1 + z) \times EW_{0}}{1 + \rho/A(z)}
\end{equation}
where EW$_{obs}$ is the observed minimum equivalent width, EW$_{0}$ is the equivalent width of the feature in the host galaxy template of  \cite{kin96}, $\rho$ is the nucleus-to-host flux ratio, $z$ is the redshift and $A(z)$ is the aperture correction. 
The lower limits are reported in Table \ref{tab:results}.

\begin{small}
\begin{deluxetable}{lcccllccccc}
\tabletypesize{\scriptsize}

\tablecaption{Spectral line parameters for the BL Lac objects}

\tablenum{2}
\label{tab:results}

\tablehead{\colhead{Object} & \colhead{$z$} & \colhead{$\alpha$} & \colhead{R} & \colhead{EW$_{min}$} & \colhead{Line ID} & \colhead{$\lambda$}  & \colhead{$z_{line}$} & \colhead{Type} & \colhead{FWHM}  & \colhead{EW (observed)} \\
\colhead{} & \colhead{} & \colhead{} & \colhead{} & \colhead{\AA} & \colhead{} & \colhead{\AA} & \colhead{} & \colhead{} & \colhead{km s$^{-1}$} & \colhead{\AA} \\
\colhead{(1)}           & \colhead{(2)} 	& \colhead{(3)} & \colhead{(4)}		& \colhead{(5)}       	& \colhead{(6)}		  & \colhead{(7)}    	&\colhead{(8)}         & \colhead{(9)}  & \colhead{(10)} 	& \colhead{(11)}\\

} 
\startdata
\object{GC 0109+224} & $>$ 0.10 & 0.85 & 14.20 & 0.11 &  &  &  &  &  &  \\

\object{PKS 0140-059} & $>$ 0.46 & 0.45 & 18.50 & 0.45 &  &  &  &  &  &  \\

\object{RBS 0337} & 0.232 & 1.30 & 16.00 & 0.35 &  &  &  &  &  &  \\

 &   &   &   &   &Ca II & 4850 & 0.233 & g & & $+0.60 \pm 0.15$ \\

 &   &   &   &   & Ca II & 4889 & 0.232 & g & & $+0.55 \pm 0.15$ \\
  &   &   &   &   & G Band & 5304 & 0.232 & g & & $+0.80 \pm 0.13$ \\
    &   &   &   &   & Mg I & 6375 & 0.233 & g & & $+1.00 \pm 0.18$ \\

\object{RBS 0499} & $>$ 0.45 & 1.00  & 18.60 & 0.65   &  &  &  &  &  &  \\
\object {PKS 0439-299} & $>$ 0.68 & 0.30 & 20.10 & 1.00 & &  &  &  &  &  \\

    &   &   &   &   & Mg II (?) & 4708 & [0.68] & i & & $+2.15 \pm 0.50$ \\

\object {RBS 0589} & $>$ 0.45 & 1.20 & 17.00 & 0.20 &  &  &  &  & & \\

\object {ZS 0506+056} & $>$ 0.15 & 0.92 & 15.30 & 0.18 &  &  &  &  & &  \\

\object {MS 0622.5-5256} & 0.513 & 1.00 & 19.30 & 0.65 &  &  &  &  & &  \\
&   &   &   &   & Ca II & 5953 & 0.513 & g & & $+1.50 \pm 0.20$ \\ 
 &   &   &   &   & Ca II & 6002 & 0.513 & g & & $+1.45 \pm 0.20$ \\
  &   &   &   &   & G Band & 6501 & 0.512 & g & & $+1.40 \pm 0.40$ \\

\object {PKS 0735+17} & $>$ 0.18 & 0.84 & 15.70 & 0.20 &  &  &  &  & &  \\  

\object {OJ-131} & 0.539 & 0.44 & 17.20 & 0.25 & &  &  &  & &  \\  
&   &   &   &   & [O II] & 5736 & 0.539 & e & $1200 \pm 150$ & $-1.15 \pm 0.30$ \\
 &   &   &   &   & H $\beta$ & 7481 & 0.539 & e & $1100 \pm 100$ & $-0.70 \pm 0.15$ \\
  &   &   &   &   & [O III] & 7703 & 0.538 & e & $1250 \pm 100$ & $-2.50 \pm 0.20$ \\

\object {RX J1022.7-0112}   & $>$ 0.30 & 1.34 & 16.60 & 0.20 &  & &  & & & \\

\object {RX J1023.9-4336} & $>$ 0.24 & 1.28 & 15.80 & 0.12 &  &  &  &  &  & \\

\object {PKS B1056-113} & $>$ 0.27 & 0.66 & 16.50 & 0.20 & & &  & & & \\

\object {1ES 1106+244} & $>$ 0.37 & 0.78 & 18.20 & 0.65 & & &  & & & \\ 

\object {RX J1117.0+2014} & 0.140 & 1.40 & 16.10 & 0.40 & & &  & & & \\ 
 &   &   &   &   & [O II] & 4245 & 0.139 & e & $1200 \pm 120$ & $-1.20 \pm 0.20$ \\
 &   &   &   &   & Ca II & 4489 & 0.140 & g &  & $+0.70 \pm 0.15$ \\
  &   &   &   &   & Ca II & 4522 & 0.140 & g &  & $+0.60 \pm 0.10$ \\
    &   &   &   &   & G Band & 4908 & 0.140 & g &  & $+0.70 \pm 0.20$ \\
      &   &   &   &   & Na I Blend & 6710 & 0.139 & g &  & $+1.10 \pm 0.40$ \\

\object {RX J1241.8-1455} & $>$ 0.32 & 1.05 & 16.80 & 0.20 & & & &  &  & \\

\object {1ES 1239+069} & $>$ 0.47 & 1.00 & 18.10 & 0.35 & & & &  &  & \\ 

\object {4C-12.46} & $>$ 0.37 & 0.43 & 17.60 & 0.30 & & & &  &  & \\

\object {OQ-240} & $>$ 0.10 & 1.00 & 14.20 & 0.15 &  &  & &  &  &  \\ 

\object {RBS 1478} & $>$ 0.50 & 0.95 & 18.60 & 0.50 & & & &  &  & \\ 

\object {OW-080} & $>$ 0.47 & 0.42 & 18.10 & 0.25 &  &  & &  &  &  \\ 

\object {OX-183} & 0.870 & 0.16 & 18.70 & 0.50 & &  & &  &  &  \\ 

&   &   &   &   & Mg II & 5236 &  0.870 & e & $4400 \pm 300$ &  $-5.80 \pm 1.60$ \\
&   &   &   &   & [O II] & 6970 & 0.870 & e & $1100 \pm 200$ & $-3.50 \pm 0.70$ \\

\object {RBS 1899} & $>$ 0.53 & 0.93 & 18.60 & 0.45 & & &  &  &  & \\  

\object {RX J2319.6+1611} & $>$ 0.97 & 0.96 & 17.20 & 0.25 & & &  &  &  & \\  
&   &   &   &   & Mg II (?) & 5527 & [0.97] & i & & $+1.50 \pm 0.30$ \\

\object {1ES 2322-409 } & $>$ 0.16 & 0.84 & 15.90 & 0.32 &  &  &  &  & & \\
\enddata

\tablecomments{Description of columns: 
(1) Object name;     
(2) average redshift (or lower limit); 
(3) power-law spectral index;
(4) R magnitude measured on the spectrum; 
(5) Minimum equivalent width detectable; 
(6) Line identification 
(7) Barycenter of the line 
(8) Redshift of the line; 
(9) Line type
\textbf{e}: emission line from BL Lac;  
\textbf{g}: absorption line from BL Lac host galaxy;
\textbf{i}: absorption line from intervening system;
(10) Full Width at half maximum;
(11) Measured equivalent width
}

\end{deluxetable}
\end{small}

\begin{small}
\begin{deluxetable}{lcclllccl}
\tabletypesize{\scriptsize}

\tablecaption{Spectral line parameters for the objects confirmed non-BL Lacs}

\tablenum{3}
\label{tab:resultsqso}

\tablehead{\colhead{Object} & \colhead{$z$} & \colhead{R} &  \colhead{Line ID} & \colhead{$\lambda$}  & \colhead{$z_{line}$} & \colhead{Type} & \colhead{FWHM}  & \colhead{EW (observed)} \\
\colhead{} & \colhead{} & \colhead{} & \colhead{} & \colhead{\AA} & \colhead{} &   \colhead{} &  \colhead{km s$^{-1}$} & \colhead{\AA} \\
\colhead{(1)}           & \colhead{(2)} 	& \colhead{(3)} & \colhead{(4)}		& \colhead{(5)}       	& \colhead{(6)}		  & \colhead{(7)}    	&\colhead{(8)}         & \colhead{(9)}\\
} 

\startdata

\object{PKS 0302-623} & 1.350 & 17.70 & & & & & &  \\
& & &  C III] & 4485 & 1.349 & e & 5500 $\pm$ 300  & $-20.00 \pm 2.00$ \\
& & & Mg II (?) & 5756 & [1.056] & i &  &  $+2.50 \pm 0.30$ \\
& & & Mg II & 6585 & 1.350 & e & 7000 $\pm$ 300 &  $-35.00 \pm 8.00$ \\

\object{PKS 0410-519} & 1.254 & 17.10 & & & & & &  \\
& & & C III] & 4300 & 1.252 & e & 5200 $\pm$ 300 & $-63.50 \pm 5.00$\\
 & & &  Mg II & 6310 & 1.253 & e & 3400 $\pm$ 300 &  $-48.00 \pm 6.00$ \\

\object{PKS 0443-387} & 0.537 &   16.80 & & & & & &  \\
 & & & Mg II & 4298 & 0.535 & e & 7400 $\pm$ 500 & $-53.50 \pm 3.00$ \\
& & &  [Ne V] & 5143 & 0.537 & e & 1100 $\pm$ 100 &  $-0.60 \pm 0.20$ \\
& & &  [Ne V] & 5268 & 0.537 & e & 1500 $\pm$ 200 &  $-2.60 \pm 0.30$ \\
& & &  [O II] & 5730 & 0.537 & e & 1200 $\pm$ 100 &  $-2.50 \pm 0.20$ \\
& & &  [Ne III] & 5946 & 0.537 & e & 2000 $\pm$ 300 &  $-5.30 \pm 0.60$ \\
& & & H $\epsilon$ & 6098 & 0.537 & e & 1200 $\pm$ 200 &  $-1.70 \pm 0.30$ \\
& & & H $\delta$ & 6304 & 0.537 & e & 1050 $\pm$ 150 &  $-0.90 \pm 0.20$ \\
& & & H $\gamma$ & 6670 & 0.537 & e & 5300 $\pm$ 400 &  $-30.00 \pm 3.00$ \\
& & & H $\beta$ & 7455 & 0.534 & e & 5300 $\pm$ 400 &  $-30.00 \pm 3.00$ \\
& & & [O III] & 7697 & 0.537 & e & 1100 $\pm$ 100 &  $-35.00 \pm 8.00$ \\

\object{PMN J0529-3555} & 0.323 & 19.20 & & & & & &  \\

 & & & Mg II & 3705 & 0.323 & e & 6400 $\pm$ 400 & $-90.30 \pm 10$    \\
& & &  Ca II & 5206 & 0.323 & g & &  $+4.60 \pm 0.70$ \\
& & &  Ca II & 5250 & 0.323 & g & &  $+4.50 \pm 0.50$ \\
& & &  G Band & 5689 & 0.323 & g & &  $+4.50 \pm 0.50$ \\
& & &  [O III] & 6560 & 0.323 & e &1300 $\pm$ 100 &  $-4.70 \pm 0.80$ \\
& & &  [O III] & 6623 & 0.323 & e & 1100 $\pm$ 200 &  $-9.80 \pm 1.20$ \\

\object{PKS 0539-543} & 1.191 & 17.60 & & & & & &  \\
  & & & C II] & 5097 & 1.191 & e & 3200 $\pm$ 300 & $-1.40 \pm
 0.30$ \\
 & & &  Mg II & 6135 & 1.191 & e & 4000 $\pm$ 400 &  $-50.00 \pm 8.00$ \\

\object{PKS 0607-605} & 1.100 & 17.30 & & & & & &  \\
 & & & C III] & 4000 & 1.095 & e & 3400 $\pm$ 300 & $-30.90 \pm 3.00$\\
& & &  Mg II & 5880 & 1.100 & e &  4000 $\pm$ 300 &  $-45.50 \pm 4.00$  \\
& & &  [O II] & 7821 & 1.100 & e &  1100 $\pm$ 100 &  $-13.50 \pm 2.00$  \\

\object{OR 103} & 1.833 & 15.70 & & & & & &  \\
 & & & Mg II (?) & 3778 & [0.349] & i &   & $+2.80 \pm 0.5$ \\
& & &  C IV & 4393 & 1.834 & e & 5200 $\pm$ 300 &  $-13.50 \pm 2.00$ \\
& & &  C III] & 5410 & 1.833 & e & 4800 $\pm$ 400 &  $-5.00 \pm 0.70$ \\

\object{PKS 1815-553} & 1.633 & 20.80 & & & & & &  \\
 & & &  C III] & 5030 & 1.634 & e & 6800 $\pm$ 600 & $-92.00 \pm 13.00$ \\ 
& & &  Mg II & 7370 & 1.632 & e & 6100 $\pm$ 400 &  $-115.00 \pm 15.00$ \\

\enddata

\tablecomments{Description of columns: 
(1) Object name;     
(2) average redshift; 
(3) R magnitude measured on the spectrum; 
(4) Line identification 
(5) Barycenter of the line 
(6) Redshift of the line; 
(7) Line type
\textbf{e}: emission line from BL Lac;  
\textbf{g}: absorption line from BL Lac host galaxy;
\textbf{i}: absorption line from intervening system;
(8) Full Width at half maximum;
(9) Measured equivalent width
}

\end{deluxetable}
\end{small}

\setcounter{figure}{2}
\section{The mean spectrum of BL Lac objects}
\label{sec:specmean}
\begin{figure*}[htbp]
  \resizebox{\hsize}{!}{\includegraphics{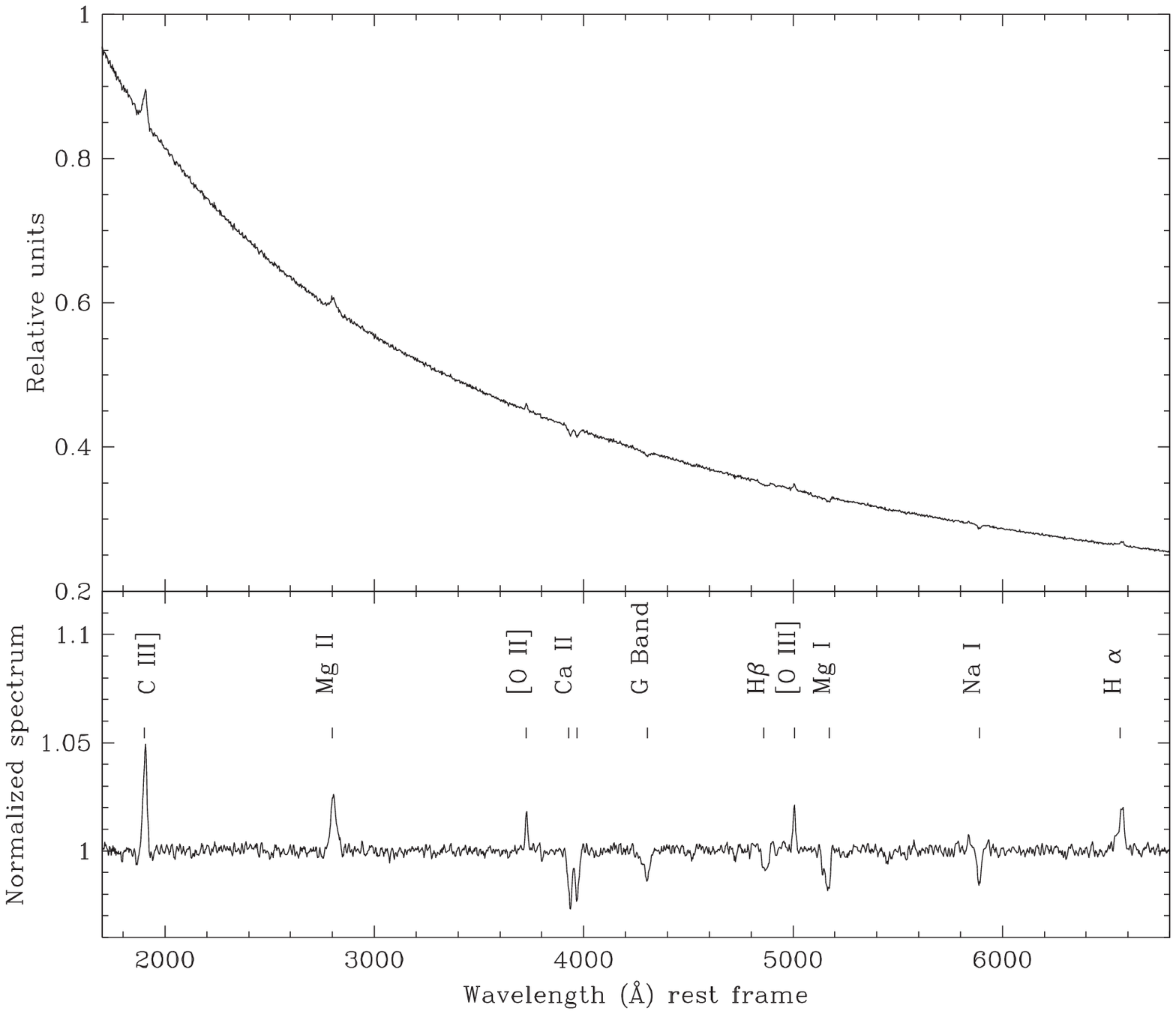}}
  \caption{Mean spectrum of BL Lac objects obtained combining the 23 objects of our campaign in which intrinsic spectral features are detected. The first panel reports the mean spectrum assuming for the continuum a power-law with index $\alpha = 0.90$ (which corresponds to the mean spectral index of the whole BL
Lac sample). In the second panel normalised spectrum is shown. }\label{fig:specdec}
\label{fig:specmean}
\end{figure*}

\begin{deluxetable}{lllllllc}
\label{tab:resavgmean}
\tabletypesize{\scriptsize}
\tablenum{4}
\tablecaption{Spectral line parameters from the BL Lac mean optical spectrum.}
\tablehead{
\colhead{Line identification}&
\colhead{$\lambda$ }&
\colhead{EW}&
\colhead{FWHM} &
\colhead{EW [min; max]} &
\colhead{FWHM [min; max] } &
\colhead{$N_{obs}$ / $N_{tot}$} &
\colhead{Line type} \\
\colhead{} & \colhead{\textrm{\AA}} & \colhead{\textrm{\AA}} & \colhead{km s$^{-1}$} & \colhead{\textrm{\AA}} & \colhead{km s$^{-1}$} &   \colhead{}  & \colhead{}\\
\colhead{(1)}           & \colhead{(2)} 	& \colhead{(3)} & \colhead{(4)}		& \colhead{(5)}       	& \colhead{(6)}		  & \colhead{(7)}  & \colhead{(8)}\\
}
\startdata
C III] & 1909 & $1.20 \pm 0.10$ & 4000 $\pm$ 350 & $[1.30;1.90]$  & [2000; 4000] & 3/4 &e \\
Mg II & 2800 & $1.00 \pm 0.20$ & 3500 $\pm$ 200 & $[0.30; 3.20]$ & [1900; 6100] & 9/13 & e \\
{}[O II] & 3727 & $0.30 \pm 0.10$ & 1300 $\pm$ 200 & $[0.10; 2.50]$& [1000; 4700]& 9/18 & e \\
Ca II  & 3934 & $0.80 \pm 0.20$ & & $[0.30; 6.10]$ & & 12/17 &g  \\
Ca II & 3968 & $0.65 \pm 0.10$  & & $[0.40; 4.20]$ & & 12/17 &g \\
G Band & 4305 & $0.75 \pm 0.10$ & & $[0.42; 4.00]$ & &12/17 &g \\
H $\beta$ & 4862 & $0.25 \pm 0.10$ & & $[0.50; 3.00]$ & & 2/14& g\\
{}[O III] & 5007 & $0.50 \pm 0.10$ & 1100 $\pm$ 200 & $[0.80; 1.47]$& [500; 2400] & 3/14 & e\\
Mg I & 5162 & $0.90 \pm 0.20$ & & $[0.26; 14.00]$ & & 3/12&g \\
Na I & 5892 & $0.40 \pm 0.10$ & & $[1.00; 3.00]$ & & 2/6&g \\
H $\alpha$ & 6563 & $0.90 \pm 0.15$ & 1300 $\pm$ 200 & * & *  &1/3 & e \\
\enddata
\tablecomments{Description of columns: 
(1) Ion identification;
(2) Rest frame wavelength;
(3) Measured equivalent width
(4) Full width at half maximum
(5) Equivalent width measurement range
(6) Full width at half maximum measurement range
(7) Ratio between number of detections and number of spectra in which the line could be measures
(8) Type of line:
\textbf{e}: emission line from BL Lac;  
\textbf{g}: absorption line from BL Lac host galaxy;
}
\end{deluxetable}

In order to investigate the overall spectral characteristics of the BL Lac objects, we constructed a mean optical spectrum combining the normalized spectra of the 23 sources with $z$ measured through intrinsic emission or absorption lines. In details, for each spectrum we first carefully removed spectral region contaminated by telluric, interstellar medium or intervening system absorption and then we corrected the spectrum to rest frame reference by adopting the measured $z$. 
We finally averaged the spectra according to their S/N ratio. The obtained mean spectrum is reported in Figure \ref{fig:specmean} and it covers the spectral range 1800 \AA $\sim$ 6700 \AA, rest frame. The first panel shows the mean spectrum assuming for the continuum a power-law with index $\alpha = 0.90$ while the normalised spectrum is reported in the bottom panel. We measure the spectral features detectable on the average spectrum reporting the line identification, equivalent width and full width at half maximum in Table \ref{tab:resavgmean}. We note that due to the large difference in BL Lacs of the ratio between the non-thermal emission and the thermal one associated to stellar component (their host galaxy) the mean BL Lac spectrum cannot be considered as representative as that of quasars. On the other hand, it illustrates emission or absorption lines that could be measured in individual objects. In particular, absorption lines from the host galaxy are always present in the spectrum but their detectability depends both on S/N ratio and on the nucleus-to-host ratio of the source (see also \ref{sec:eelt}). Conversely, emission lines could be missing.

\section{Discussion and conclusions}
\label{sec:discussion}
\subsection{Summary on the campaign}
\label{subsec:sample}

According to the 11th Veron-Cetty catalog of AGN objects (Veron 2003) at the time of the scheduling of our campaign in 2003, 797 sources were known as candidate or confirmed BL Lacs and 313 objects of the list were left without a reliable measure of their redshift.
Considering only the sources observable from the southern sky ($\delta < 15$) a total of 169 BL Lacs objects remains of unknown redshift.
From this set, we selected objects with $V < 21$ in order to secure an acceptable S/N ratio spectrum for each source. This additional constraint allowed us to select $\sim$ 155 observable sources from the Paranal with VLT+FORS.
We obtained time for the observation of 88 sources selected combining the 155 objects from the 11th Veron catalog with further 44 sources from Padovani and Giommi 1995a.
Thanks to the high quality optical spectra, we confirmed as bona-fide BL Lacs 69 sources while 15 have been classified as QSO or FSRQ. The remaining 4 objects are stars.
For the sources that have been classified as bona-fide BL Lac we were able to measured a spectroscopic redshift for 23 of them. In particular, in 12 sources we clearly detected the absorption features from their host galaxy and in 15 objects emission lines are observable. Broad emission lines (such as Mg II or C III]) are detected in only 10 BL Lacs. We measured the luminosity of Mg II emission lines in all the sources in which the line is detected. The mean value is $\log \textrm{L(Mg II)}  \sim 42.40 \pm 0.30$, where $\textrm{L (Mg II)}$ is in erg s$^{-1}$ . This value is lower of an order of magnitude of that of quasars which is   $\log \textrm{L(Mg II)} \sim 43.60 \pm 0.45$ (\cite{shen11}) but the dispersions are rather larger.
For the remaining 46 sources we inferred a redshift lower limit (see Section \ref{sec:specdec}) . The full dataset, with spectra and line identification, is available at the website \textit{http://www.oapd.inaf.it/zbllac/}.

\subsection{Optical beaming factor}

In the radio emission, the relativistic beaming is manifest in BL Lac objects because of the frequent  appearance of jets with superluminal velocities.
This is required in high energy emissions (in $\gamma$-rays) by the short time scale of variability,
which can be accommodated in a standard SSC model (e.g. \citep{maraschi03}) only invoking large Doppler beaming factors. 
Following the unified model of Active Galatic Nuclei we assume that the optical emission arising from the nuclear component is basically the superposition of a thermal emission of luminosity $L_{th}$ associated with a disk structure and a non thermal one $L_{nth}$, which dominates the output power in BL Lac objects. Since the former component is isotropic and the latter supposedly is beamed, we can define as beaming factor (see e.g. \cite{farina12}, \cite{landoni12})
\begin{equation}
\delta = \frac{L_{tot}}{L_{th}}
\end{equation}
where $L_{tot}$ = $L_{th}$ + $L_{nth}$.

\begin{figure}[htbp]
    \centering

  \resizebox{\hsize}{!}{\includegraphics{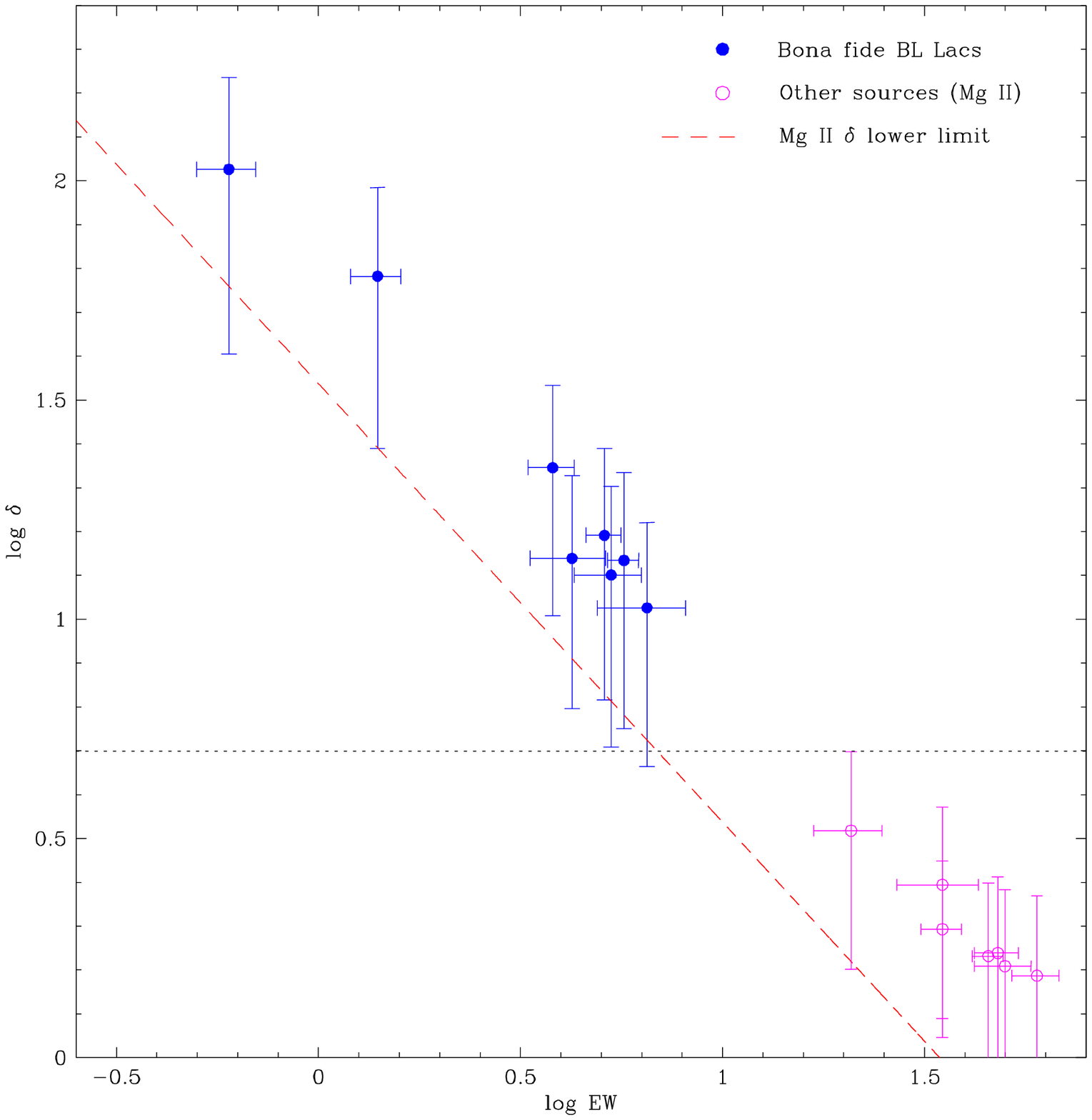}}
  \caption{Optical beaming factor for the sources observed in the campaign. Red dashed line indicates the lower limit value calculated through Equation 4 assuming $z$ = 0.70. Filled blue circle are the $\delta$ of bona-fide BL Lac objects with Mg II emission line  while the open magenta circle are the beaming factor for other intermediate sources (classified as confirmed non-BL Lacs). Dotted line delimit the area of the EW $- \delta$ plane where the source are of intermediate nature between pure QSO ($\delta \sim $1)  and BL Lacs ($\delta \gtrsim$ 4).}
  \label{fig:overallbf}
\end{figure}

    \setcounter{figure}{3}
We further assume that the thermal components of BL Lac objects can be described as that of quasars. The large collection of Type I AGN spectra available in the SDSS allows to find relations between the broad line luminosity $L_{line}$ and the contiguous (thermal) luminosity for Type I  quasars.  They are generally expressed in the form

\begin{equation}
\log \frac{\lambda L_{\lambda} (\lambda_{ref})}{L_{line}} = K
\end{equation}
For simplicity, we refer only to Mg II relation since it is the tightest one ($K = 1.91 \pm 0.26$, $\lambda_{ref} = 3000$, see \cite{decarli11, shaw12}). 
Taking into account the contribution of the cosmological redshift and combining equation (2) and (3) with the usual defintion of the equivalent width , one has
\begin{equation}
\delta = (1+z) \times \frac{\lambda_{ref}}{10^{K} \times EW}
\end{equation}
$\delta$ is obviously equal to $\sim$1, within the dispersion, in the case Type I quasar.
When a broad Mg II line is observed in a BL Lac object Eq. (4) gives an estimate of the beaming. If no emission line is detected one can still use Eq. (4)
and from the upper limit on the equivalent width deduce a lower limit on $\delta$. We suppose that the redshift is such that the relation can be applied because the Mg II emission line is within band of observation and we also consider the positive 1-$\sigma$ dispersion on the relation, which yields $K = 2.17$ . In particular, for featureless BL Lacs we assume $z \sim 0.7$ while for the BL Lac in which the redshift is known we use the actual value for the redshift. The wide-spread criterion of defining a BL Lac a Type I AGN with line $EW < 5$, which was adopted also in this paper, therefore translates through Eq. (4) in $\delta$ $\gtrsim 4$.
\\
In Figure \ref{fig:overallbf} we show the lower limit relation expressed in Eq. (4) for the beaming factor of BL Lacs with no broad emission lines (dashed line) and the actual value of $\delta$ for the sources in which the broad Mg II emission line is detected (filled blue circle).
We note that $\sim$ 7 BL Lac candidates considered in our sample have spectral properties that do not completely satisfy the criterion for BL Lac classification. On the other hand, these objects differ also from the typical spectrum of Type I AGN. Interesting enough they have their optical beaming factor (open magenta circle of Figure \ref{fig:overallbf}) much larger of that of quasar, where $\delta$ should be $\sim$ 1, suggesting a substantial non-thermal emission from the nucleus which indicates that the objects are indeed of intermediate nature between BL Lac and quasars.  

\subsection{Mg II intervening absorption systems}
\label{sec:MgIIintervening}
Since BL Lac objects are extragalactic sources it is reasonable to expect that their spectra could present absorption features from intervening systems. In particular, \cite{stocke97} claimed that a strong excess of a factor $\sim$ 4 on the number of Mg II intervening system is apparent in the spectra of BL Lac objects, with respect to those of quasars.
\\
In our sample there are 7 confirmed BL Lacs objects in which absorption systems are detected. These absorptions are interpreted as Mg II intervening systems on our line of sight toward the BL Lac.
In particular, the total number of absorbing systems is $N_{obs}$ = 8, partly associated with BL Lacs with unknown redshift. The most natural interpretation is that the features are due to intervening MgII at $\lambda_{0} =  2800$.
The features associated to intervening Mg II systems are observable in our sample in the redshift interval $z_{min} \sim 0.5$ and $z_{max} \sim 1.5$ .

The total number of BL Lacs we consider is $n_t$ = 69 (23 with redshift measured and 46 lower limits on $z$) and those which have $z < z_{min}$ are 13.
Therefore, excluding sources with $z < z_{min}$ the number of BL Lacs which can contribute to the observed absorptions is at most 56.
On the other hand the minimum number which for sure contributes to the statistics of the absorbers is 8, which are the observed systems associated to BL Lacs in our sample, plus the number of BL Lacs without intervening absorptions with $z>z_{min}$ that are 20.
Following \cite{zhu12}, who consider QSOs in the SDSS, for an arbitrary pointing direction the normalised mean number of MgII absorbers with EW $\gtrsim$ $1$ $\AA$ is $n_Q \sim 0.40$ in the redshift range given above.
Within these assumptions we expect that the number of Mg II absorption systems is $11 \textless$  $N_{exp}\textless 22$.
Since in our sample we observe $N_{obs} = 8$ intervening systems we do not confirm the suggestion of \cite{stocke97} and we conclude that absorbers related to BL Lacs are not more numerous than those associated to QSOs. 
Moreover, if we assume that the distribution of Mg II intervening system for BL Lac is not different from those of quasars, the number of observed absorption system could suggest that the BL Lacs without $z$ in our sample should have low value for their redshift, in order to comply with the discussed statistical properties for intervening system.

\subsection{Future perspective: BL Lac spectroscopy in the ELT era}
\label{sec:eelt}
In the spectrum of a BL Lac object the absorption lines associated to the stellar components of the host galaxy are always present. On the other hand, their detectability  depends both on the S/N ratio and on the nucleus to host ratio (N/H) of the source (see e. g. \cite{sba05}).
With the state of the art 8-mt class telescopes and modern instrumentation it is possible to obtain spectra with S/N about 100 and R $\sim$ 2000  for object of V $\sim$ 17 (assuming exposure time of $\sim$ 1 hour). For sources with $V \gtrsim 20$ the situation is even worse since the S/N typically drops below 50. 
This limits the capability to detect the host galaxy features either for the sources with high intrinsic N/H ratio and for faint objects. 
For this reason, in order to significantly improve these limits, a future 40-mt class telescope like the European Extreme Large Telescope (E-ELT) is mandatory to investigate these kind of objects.
\\
\\
\setcounter{figure}{4}
\begin{figure*}[htbp]
\centering

  \resizebox{4.5in}{!}{\includegraphics{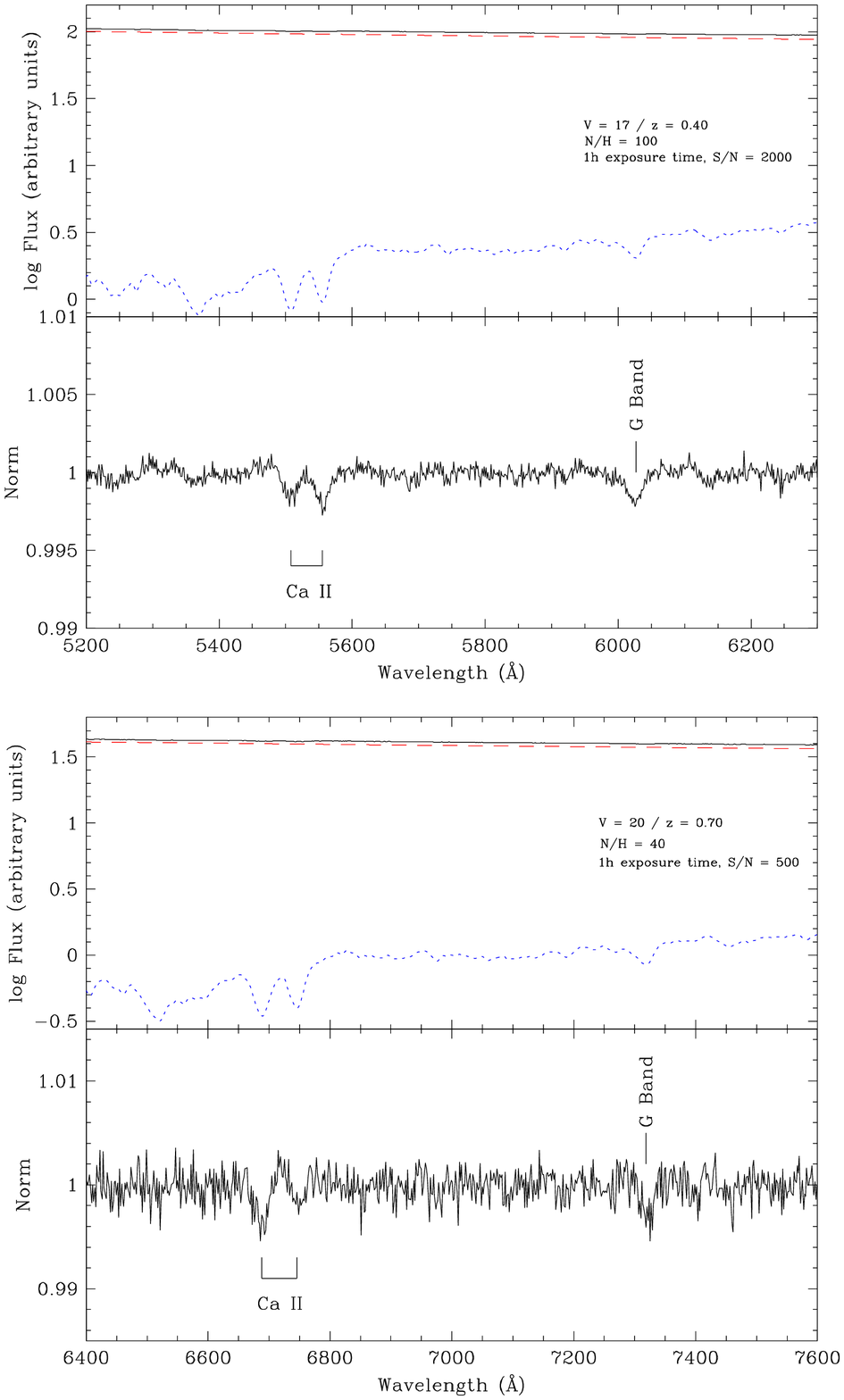}}
  \caption{First panel: simulation of a  BL Lac spectrum with V = 17, S/N = 2000, z = 0.40 and N/H = 100.  Aperture correction is 0.35 . Second panel: simulation for a BL Lac spectrum with V = 20, S/N = 500, N/H = 40, z = 0.70 and aperture correction of 0.50. For both cases the exposure time is 3600 s and the adopted slit is  a 1$^{\prime\prime}$ x 1$^{\prime\prime}$. The efficiency of the spectrograph is considered similar to the VLT-FORS2. The solid black line is the BL Lac overall spectrum while dashed line is the power law associated to the non thermal emission. Blue dotted line is the superposed host galaxy template spectrum.  }
  \label{fig:f_eelt}
\end{figure*}
As example of this capability we report here a simulation of the expected spectrum of a BL Lac object observed by E-ELT. Assuming that the BL Lac spectrum is the sum of a non thermal emission, modelled with a power law, and a stellar component described by a giant elliptical galaxy \citep{kin96} we performed the simulation of two cases. We consider an exposure time of 3600 s and a spectrograph with a dispersion of $\sim$ 1.0 $\AA$ / pixel, with an efficiency similar to the VLT-FORS2. We take into account a standard galaxy characterised  by $M_r = -23$ and $R_{eff} = 10$ kpc. We adopted an aperture correction for a slit of 1$^{\prime\prime}$ x 1$^{\prime\prime}$, similarly to the aperture effect correction fully described in Paper II.  Results are presented in Figure \ref{fig:f_eelt} . In the first case (top panel), the E-ELT can obtain a spectrum with S/N $\sim$ 2000 for an object with V = 17 and z = 0.40. The features associated to the host galaxy are clearly detectable with an intrinsic N/H up to 100 in the band 5200 - 6200 $\AA$.
The second simulation (bottom panel) is carried out assuming V = 20 and z = 0.70. In this case, the S/N ratio of the spectrum is $\sim$ 500 and the intrinsic N/H ratio limit reached is $\sim$ 40. We note that the use of the E-ELT allows to increase the S/N ratio of the spectra of a factor $\sim$ 20 for the first case and $\sim$ 10 for the latter respect to typical S/N ratios reached with the VLT.
In the light of these results, we are confident that the advent of 40-mt telescope instrumentation can significantly decrease the fraction of BL Lacs that are still left without $z$.

\acknowledgments
We acknowledge useful conversations with E. P. Farina and A. Veronesi

\newpage
\clearpage
\onecolumn
\setcounter{figure}{0}

\begin{figure}[htbp]
  \resizebox{6.25in}{!}{\includegraphics{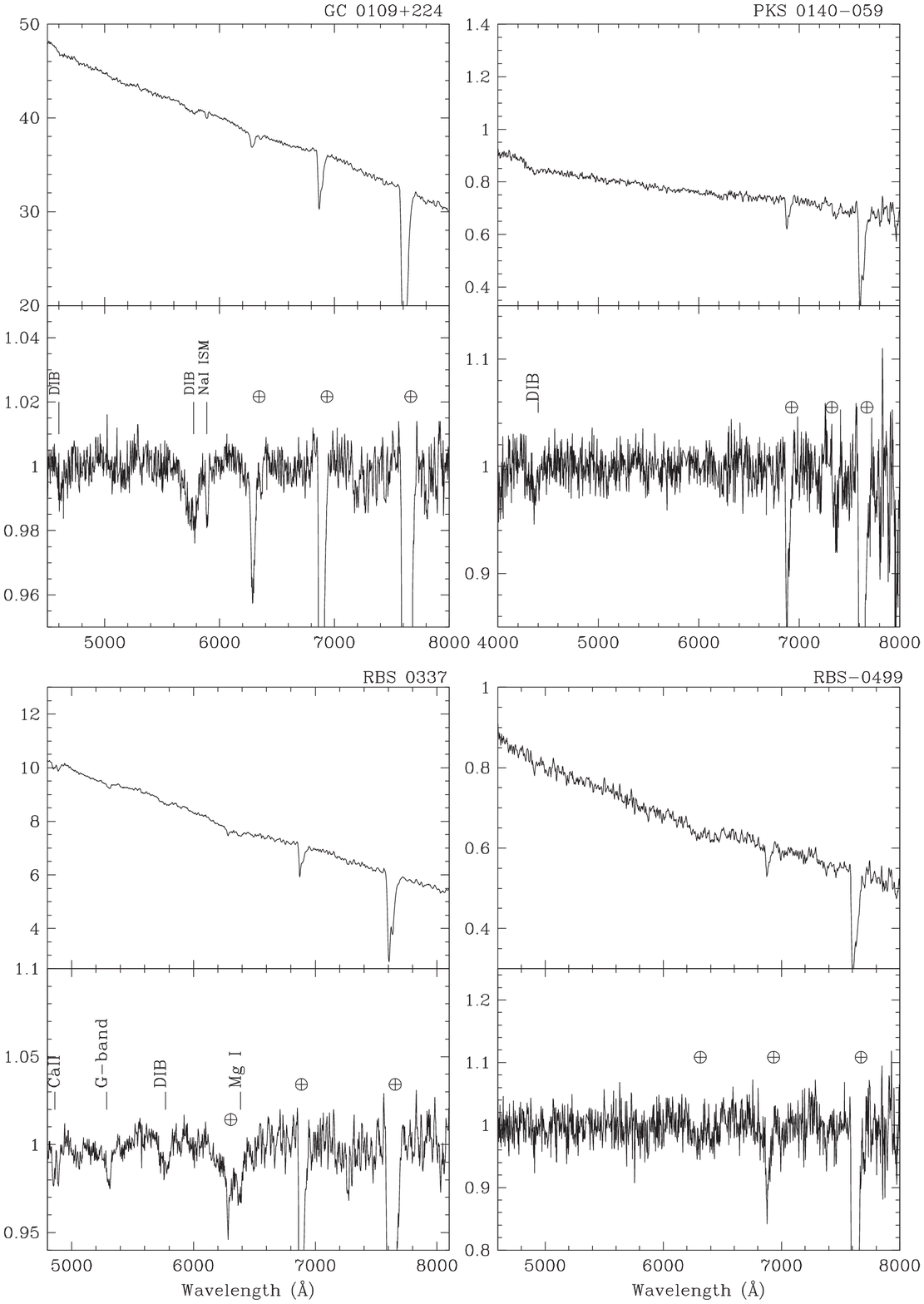}}
  \caption{Spectra of  BLLs sources. Top panel: flux calibrated spectra. Normalized spectra on bottom. Telluric bands are indicated by $\oplus$, spectral lines are marked by
   line identification, absorption features from
   interstellar medium of our galaxy are labeled by ISM, diffuse interstellar 
   bands by DIB. The flux density is in units of 10$^{-16}$ erg cm$^{-2}$ s$^{-1}$ \AA$^{-1}$.  
   }\label{fig:spectra}
\label{fig:spec}
\end{figure}

\setcounter{figure}{0}
\begin{figure}[htbp]
  \resizebox{\hsize}{!}{\includegraphics{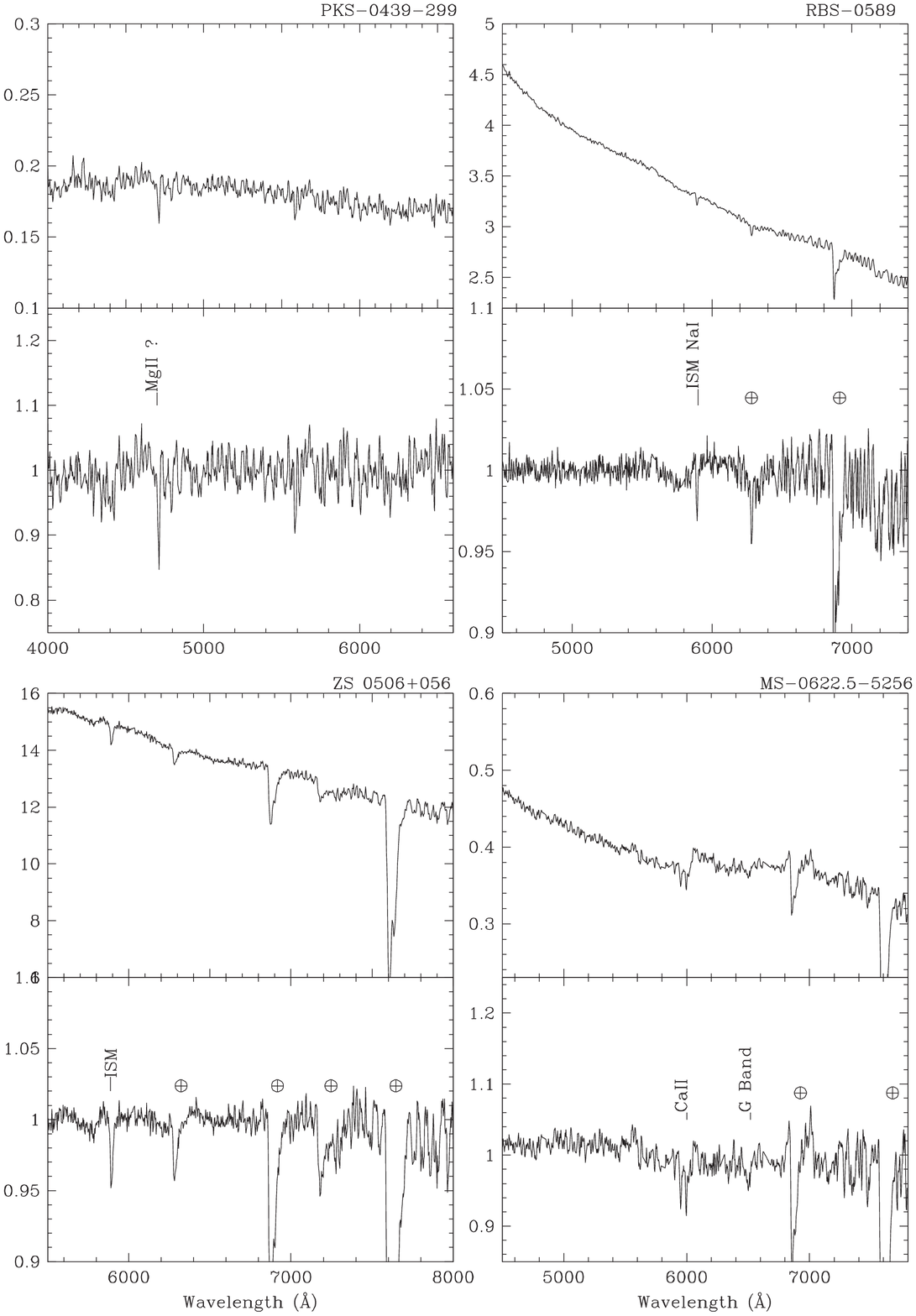}}
  \caption{continued}

\end{figure}

\setcounter{figure}{0}
\begin{figure}[htbp]

  \resizebox{\hsize}{!}{\includegraphics{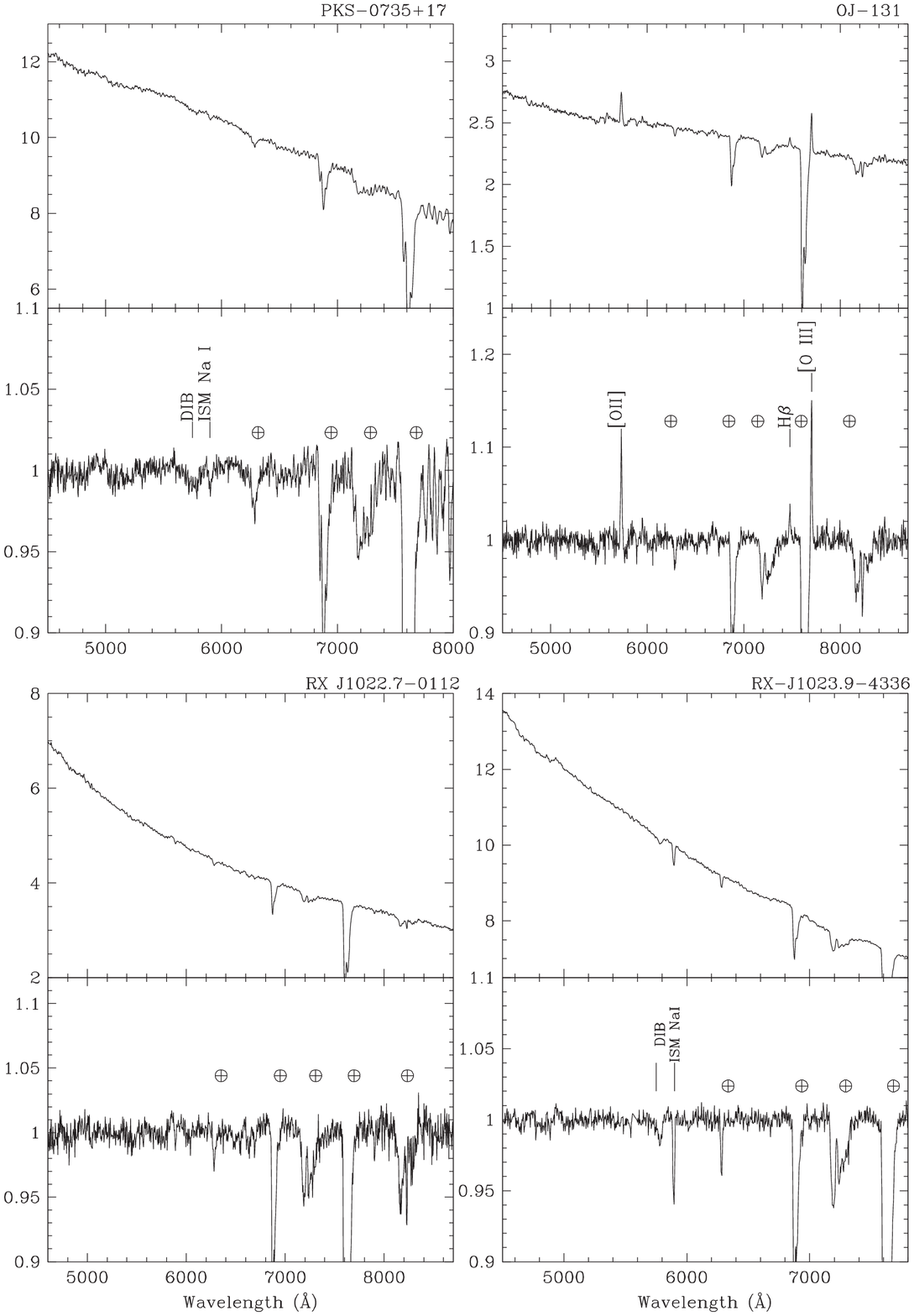}}
  \caption{continued}

\end{figure}

\setcounter{figure}{0}
\begin{figure}[htbp]
  \resizebox{\hsize}{!}{\includegraphics{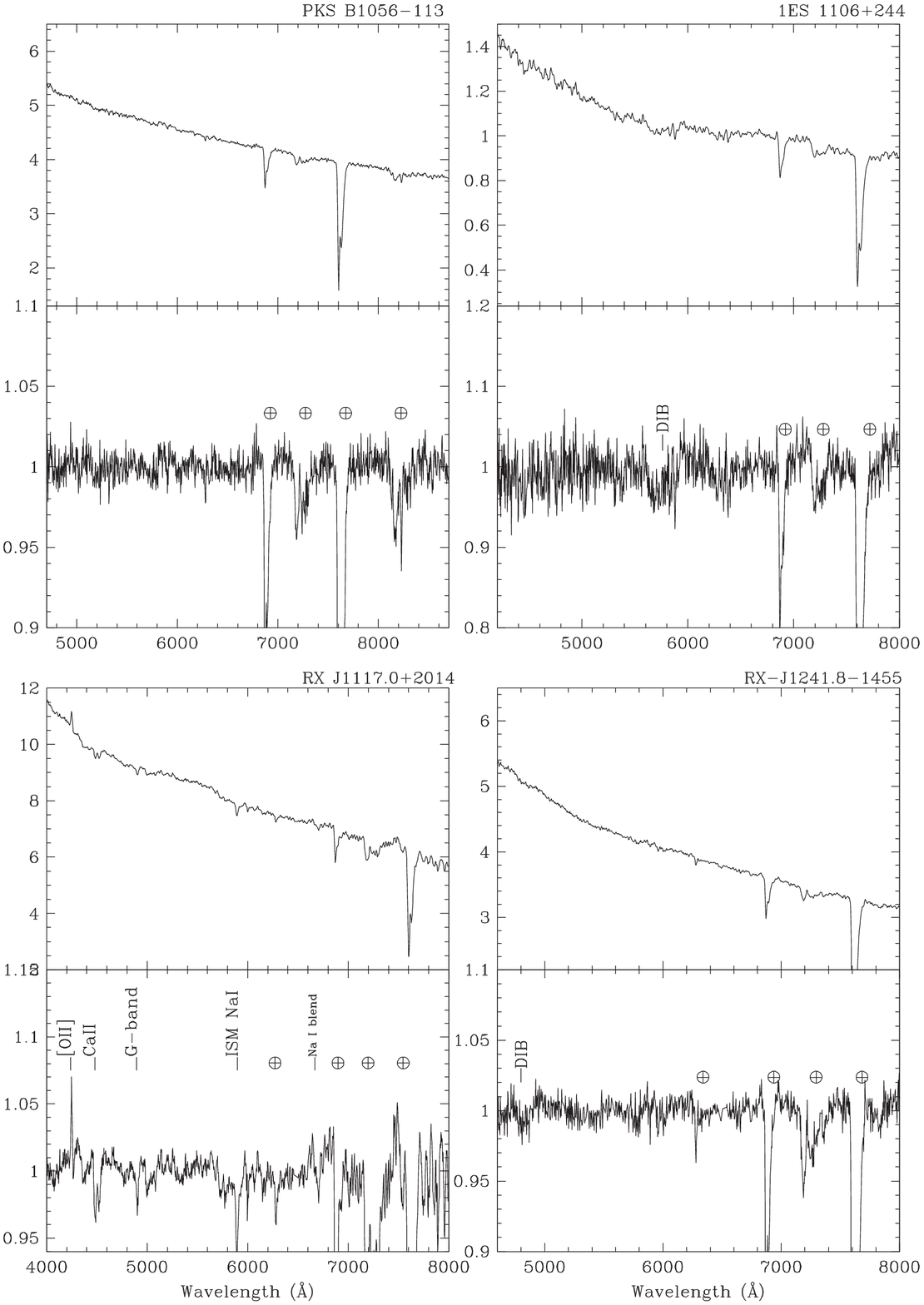}}
  \caption{continued}

\end{figure}

\setcounter{figure}{0}
\begin{figure}[htbp]

  \resizebox{\hsize}{!}{\includegraphics{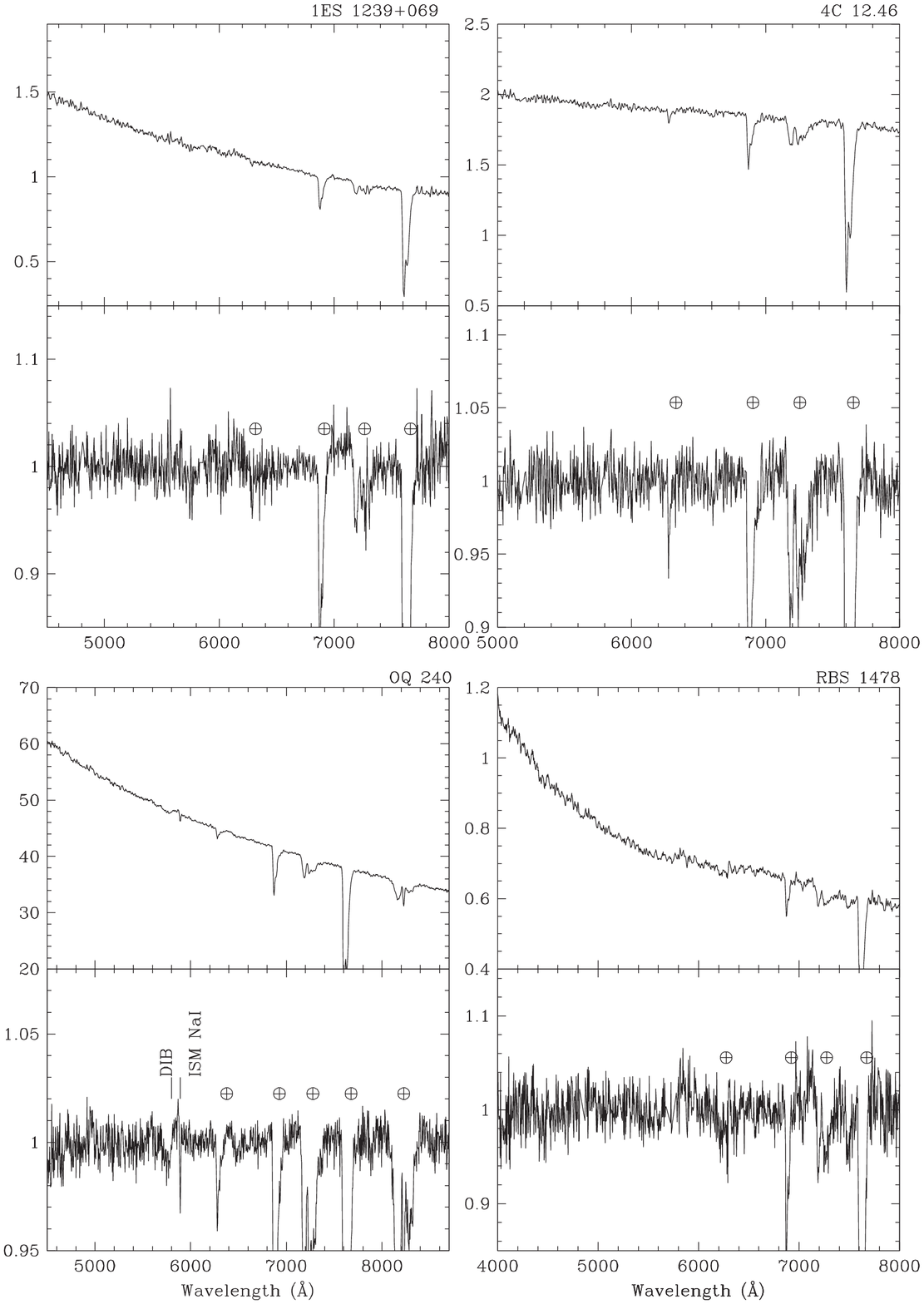}}
  \caption{continued}

\end{figure}

\setcounter{figure}{0}
\begin{figure}[htbp]
  \resizebox{\hsize}{!}{\includegraphics{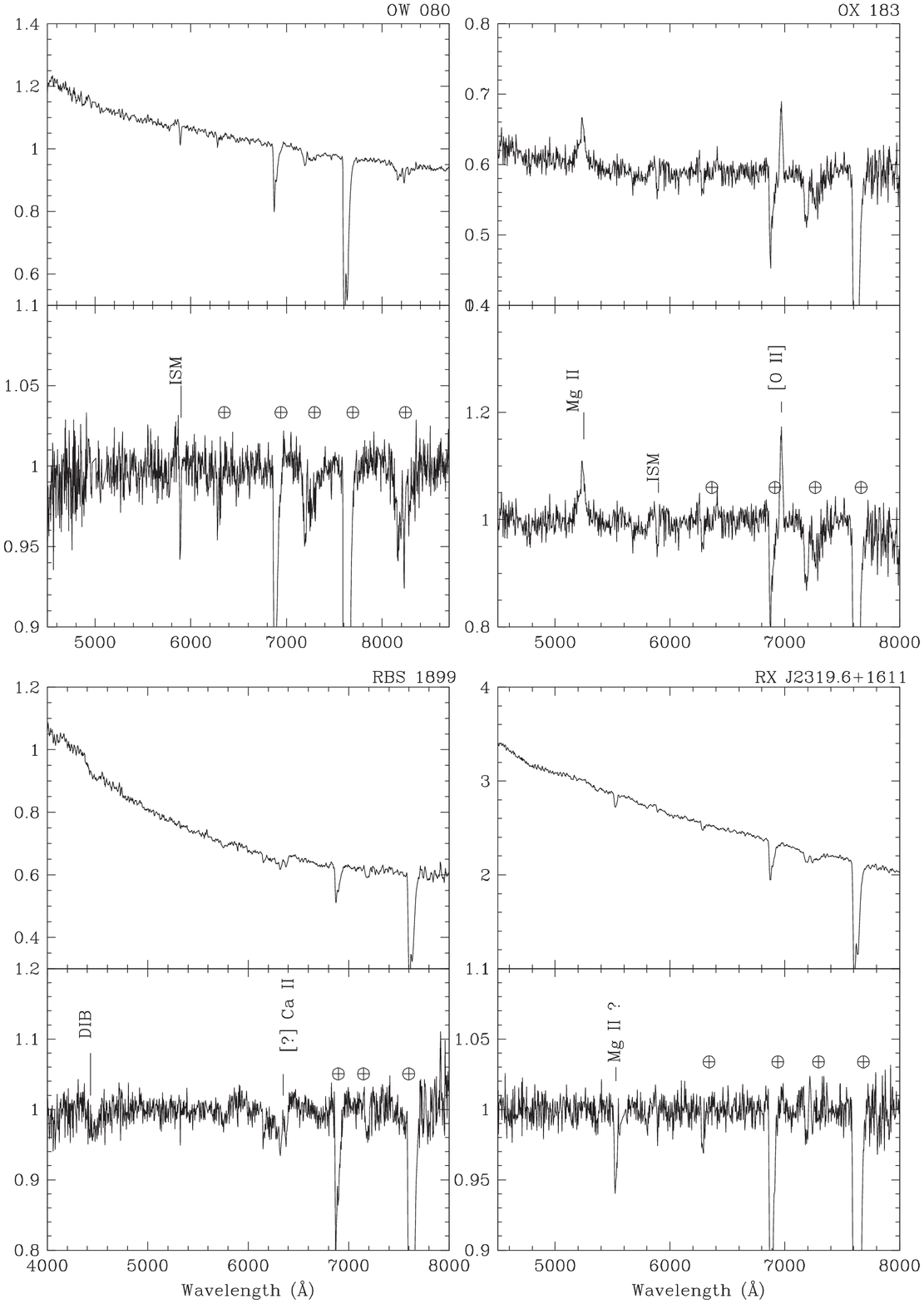}}
  \caption{continued}

\end{figure}

\setcounter{figure}{0}
\begin{figure}[htbp]
  \resizebox{\hsize}{!}{\includegraphics{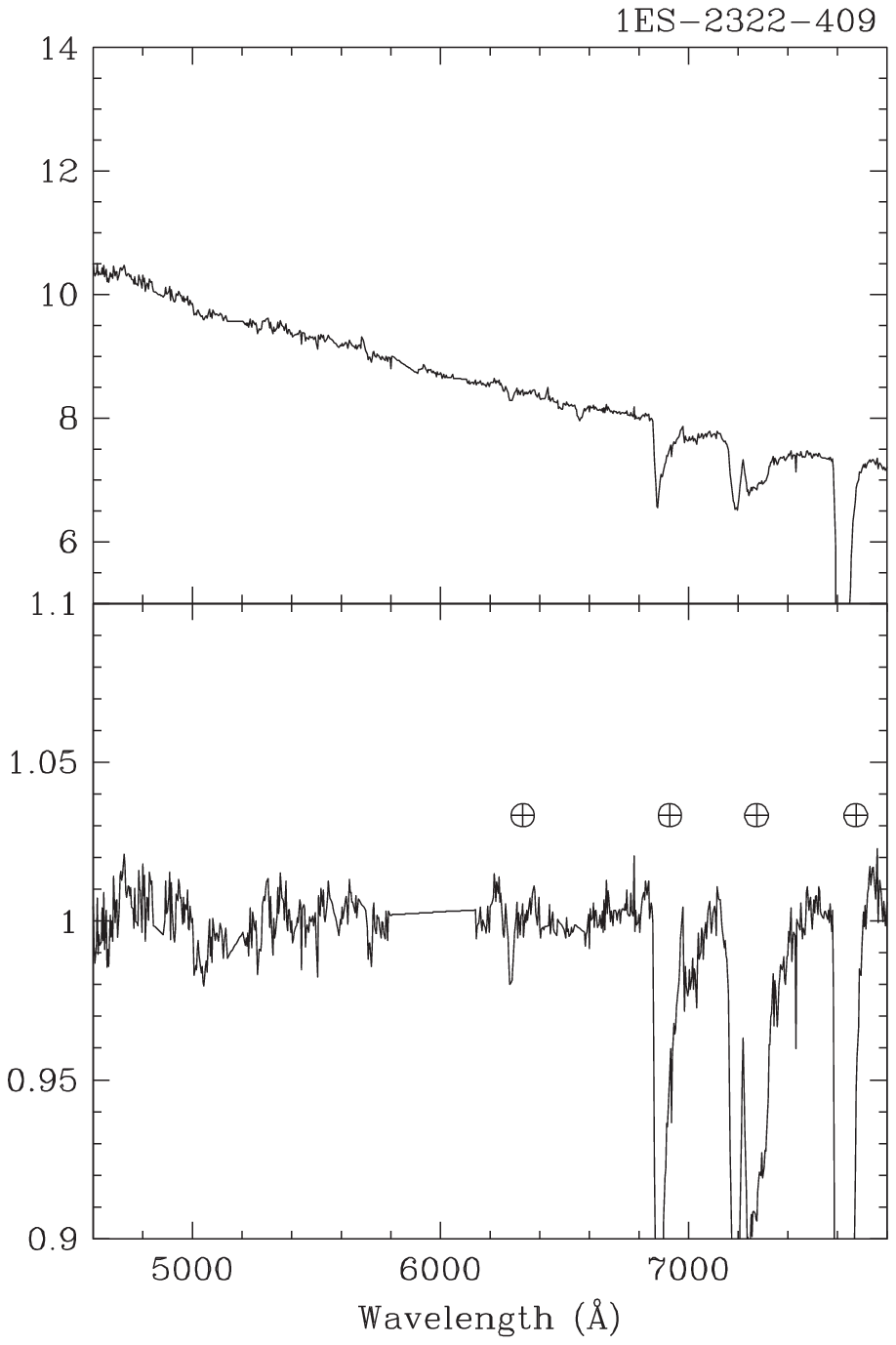}}
  \caption{continued}

\end{figure}

\setcounter{figure}{1}
\begin{figure}[htbp]
  \resizebox{6.0in}{!}{\includegraphics{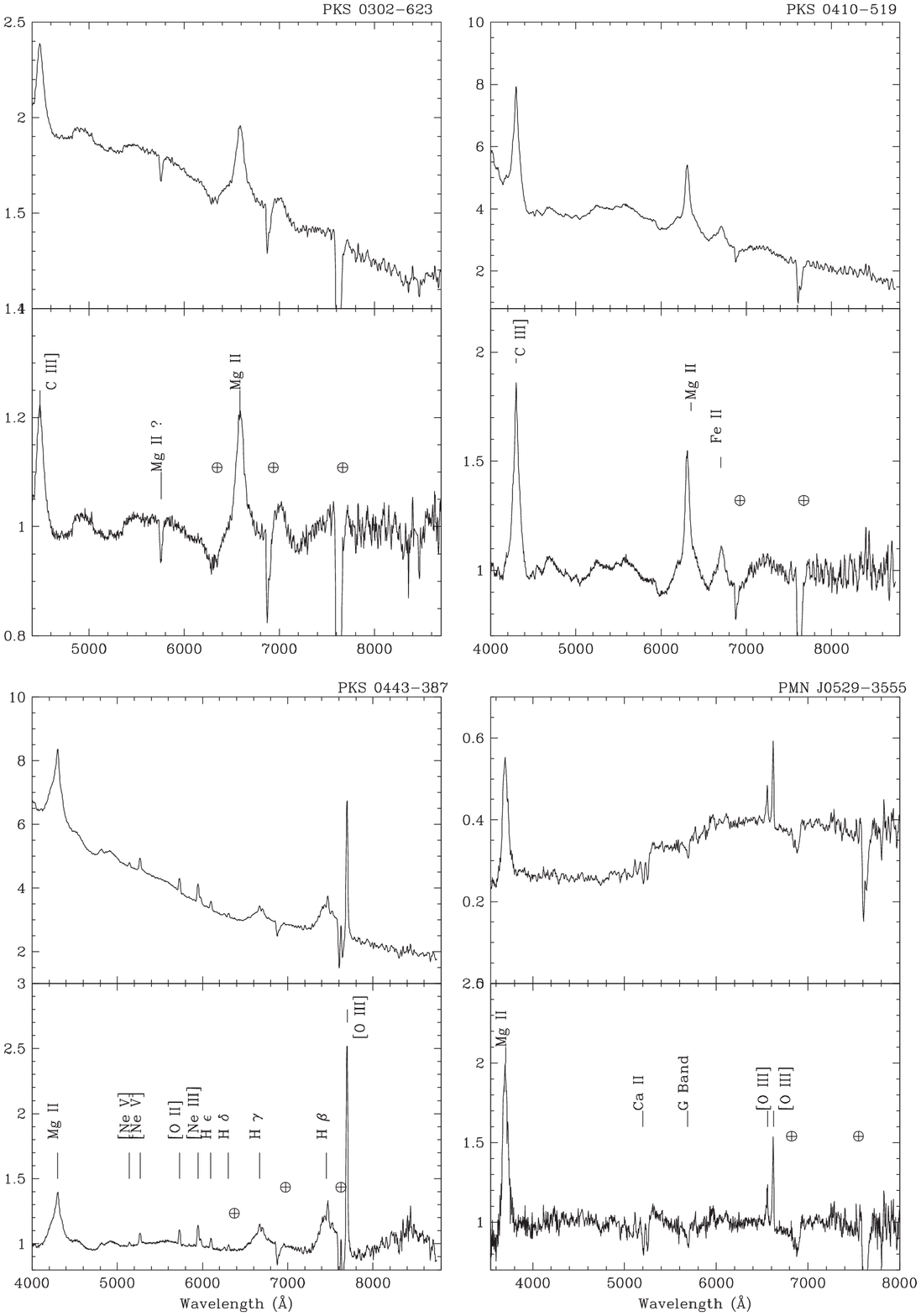}}
  \caption{Spectra of QSOs or confirmed non-BLLs sources. Top panel: flux calibrated spectra. Normalized spectra on bottom. Telluric bands are indicated by $\oplus$, spectral lines are marked by
   line identification, absorption features from
   interstellar medium of our galaxy are labeled by ISM, diffuse interstellar 
   bands by DIB. The flux density is in units of 10$^{-16}$ erg cm$^{-2}$ s$^{-1}$ \AA$^{-1}$. 
   }
   \label{fig:spectraQSO}

\end{figure}

\setcounter{figure}{1}
\begin{figure}[htbp]
  \resizebox{6.2in}{!}{\includegraphics{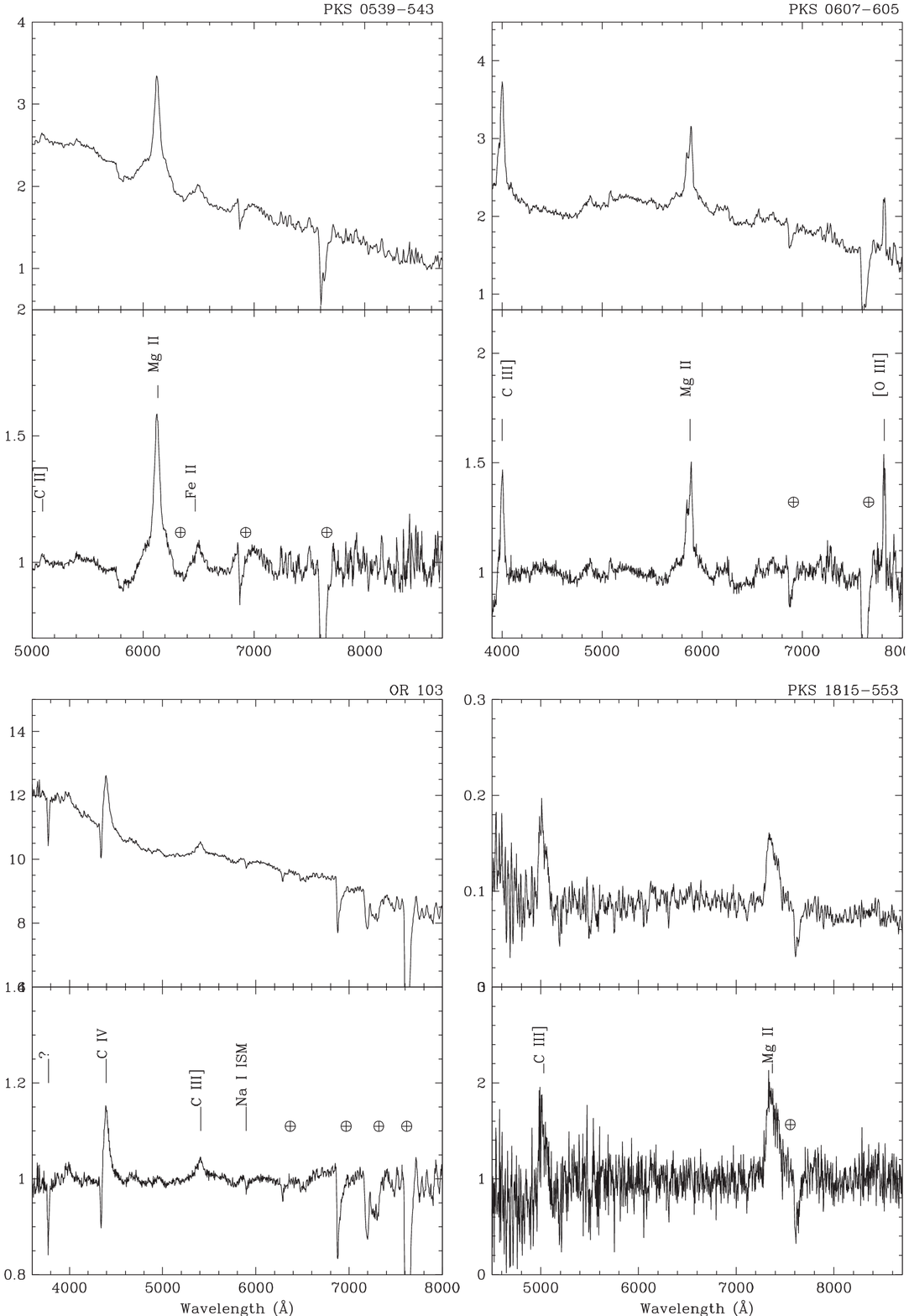}}
  \caption{continued}

\end{figure}

\newpage

\appendix
\twocolumn
\section{Notes on individual sources.}
Adopting the procedure fully described in Paper II and III we decompose the observed spectrum of objects in which the host galaxy absorption features are detected. Briefly the spectral decomposition is obtained assuming that the observed spectrum is a superposition of a non thermal component, described by a single power law, and a thermal emission from a model of a giant elliptical galaxy.
\setcounter{figure}{5}
\begin{figure}[htbp]
    \centering
  \resizebox{\hsize}{!}{\includegraphics{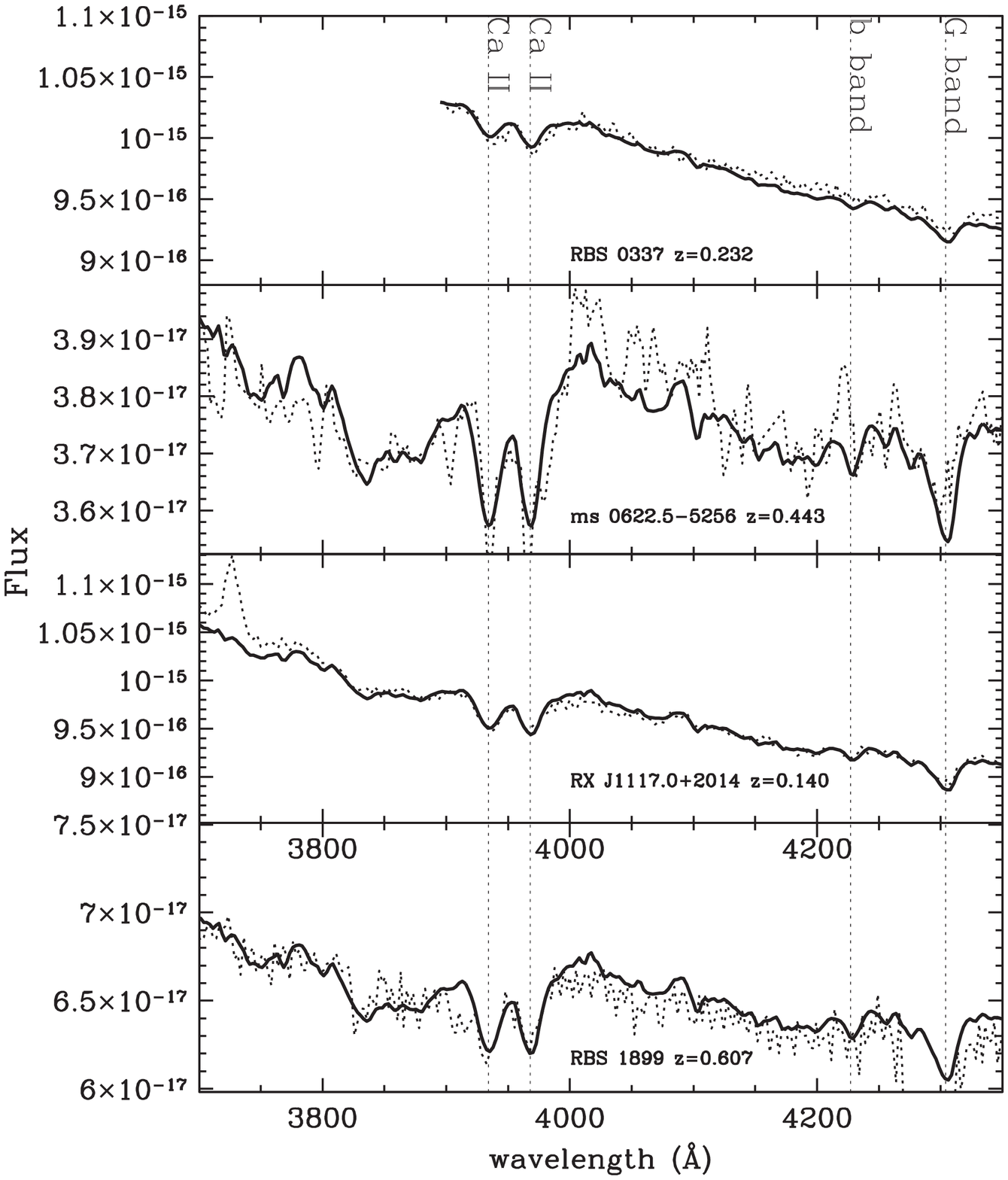}}
  \caption{Spectral decomposition (rest frame) for the objects RBS 0337, MS 0622.5-5256, RX J1117.0+2014 and RBS 1899. The solid line shows the fitted spectrum while dots is the observed one.}\label{fig:specdec}
\label{fig:specdec}
\end{figure}

\paragraph{RBS 0337} This BL Lac source has been observed in the $\gamma$-rays band by FERMI \citep{abdo10} and it has been detected in the MIR/NIR bands in the 2MASS observation of BL Lac objects II project \citep{mao11}. 
Since no imaging studies of this object are available, we inferred from the spectral decomposition presented in Figure \ref{fig:specdec} that the absolute magnitude of the nucleus and host galaxy are respectively $M_{R}^{Nuc}$ = -24.30 and $M_{R}^{Host}$ = -23.80.

\paragraph{MS 0622.5-5256} is a well known BL Lac object (see e.g. \citet{padovani95}, \citet{stocke00}) that has been studied by \citet{sba05} who reports a photometric redshift of 0.41 through the detection of the host galaxy from HST snapshot images. 
Moreover, further spectroscopical studies (\citet{sba06}) report a lower limit of $z > 0.49$ that is obviously consistent with the measurement of the redshift obtained in this work.
From the spectral decomposition obtained (see Figure \ref{fig:specdec}) we measure that the absolute magnitude of the nucleus and host galaxy are respectively $M_{R}^{Nuc}$ = -23.30 and $M_{R}^{Host}$ = -22.70 which are comparable to those measured from HST imaging \cite{sba05}.
\\

\paragraph{OJ 131} This BL Lac object has been studied in detail in the optical band by \cite{fal2000} who suggests a lower limit of z $> 0.5$ from the non detection of the host galaxy, assumed to be a standard luminous elliptical ($M_{R} = -23.80$).
The spectrum of the source presented in this paper does not show detectable spectral features deriving from the host galaxy. Our new redshift estimate is carried out  through weak emission lines arising from the nucleus itself.

\paragraph{RX J1117.0+2014} This is a BL Lac source studied in a large spectral range from $\gamma$-rays with FERMI \citep{abdo10} to the radio band \citep{becker91}. In 2001, this BL Lac has been proposed as a TeV candidate source \citep{costamante02} but up to now no VHE observation are available for this source. 
Moreover, \cite{nil07} measured the apparent magnitude of the nucleus and its host galaxy through imaging decomposition obtaining that $m_{R}^{Host}$ = 16.31 $\pm$ 0.31.
We decomposed the observed spectrum (see Figure \ref{fig:specdec}) in order to derive the absolute magnitude of the nucleus and its host galaxy that are respectively $M_{R}^{Nuc}$ = -22.70 and $M_{R}^{Host}$ = -21.50.

\paragraph{OQ+240} is a bright BL Lac object belonging to Padovani and Giommi catalog that has been observed in $\gamma-$rays from GeV up to TeV band by MAGIC and VERITAS (\cite{berger11}, \cite{maj11}). Recent study of this source in $\gamma-$rays carried out by \cite{prand10}, based on the absorption on Very High Energy photons by the Extragalactic Background Light, suggested a redshift of $0.24 \pm 0.05$ that is consistent with the lower limit derived in this paper. Moreover, adopting a similar procedure, \cite{yang10} propose an upper limit for $z$ of 1.19 for this object, well consistent with previous results.

\paragraph{OX 183} is a well kwown BL Lac object (see e.g. \cite{rector2001}) that has been studied through Hubble Space Telescope imaging by \cite{sba05} who suggested a lower limit to its redshift of 0.76 that it is obviously consistent with the $z$ derived from Mg II $\lambda 5236$ and [O II] $\lambda 6970$ emission lines reported in this work.

\paragraph{RBS 1899} Tentatively, there seem to be a Ca II break $\lambda \lambda$ 6322 6377 (EW $-1.40 \pm 0.30$, $-1.00 \pm 0.3$) in the region around $\sim$ 6300 $\textrm{\AA}$ that is contaminated by atmospheric absorption, yielding a z = 0.607. Our spectral decomposition (see Figure \ref{fig:specdec}) suggests that $M_{R}^{Nuc}$ = -24.60 and $M_{R}^{Host}$ = -23.40.

\end{document}